\documentclass[prd,aps,twocolumn]{revtex4}
\usepackage[latin1]{inputenc}
\usepackage[english]{babel}
\usepackage{graphicx}
\usepackage{color}
\usepackage{amsmath}
\usepackage{amssymb}
\usepackage{epstopdf}
\usepackage{hyperref}
\usepackage{amssymb}
\usepackage{graphicx}
\DeclareGraphicsExtensions{.pdf,.png,.jpg}
\usepackage{setspace}
\usepackage{natbib}
\usepackage{xcolor}
\usepackage{mathrsfs}
\usepackage{cancel}
\usepackage{graphics}
\usepackage{soul}
\setstcolor{red}
\newcommand{\dpar}[2]{\frac{\partial #1}{\partial #2}}
\newcommand{\ddpar}[2]{\frac{\partial^2 #1}{\partial #2^2}}

\begin{document}
\title{\bf Quasilocal Smarr relations for static black holes}
\author{F. D. Villalba, and P. Bargue\~no}
\affiliation{Departamento de F\'{\i}sica,
Universidad de los Andes, Apartado A\'ereo {\it 4976}, Bogot\'a, Distrito Capital, Colombia (fd.villalba10@uniandes.edu.co)}

\begin{abstract}
Generalized Smarr relations in terms of quasilocal variables are obtained for Schwarzschild and Reissner-Nordstr\"om black holes. The approach is based on gravitational path integrals with finite boundaries on which, following Brown and York, thermodynamic variables are identified through a Hamilton-Jacobi analysis of the action. The resulting expressions allow us to construct the relation between the quasilocal energy obtained in this setting and the Komar and Misner-Sharp energies, which are regarded as thermodynamical internal energy in other approaches. The quasilocal Smarr relation is obtained through scaling arguments, and terms evaluated in the external boundary and the horizon are present. By considering some properties of the metric, it is shown that this quasilocal Smarr relation can be regarded as a thermodynamical realization of Einstein equations. The approach is suitable to be generalized to any spherically symmetric metric.
\end{abstract}
\maketitle

\section{Introduction}\label{intro}
\par Gravity and thermodynamics seem to be connected at a very deep level. The study of black holes has shown that General Relativity implies relations between quantities defined at the event horizon that are analogous to the laws of thermodynamics \cite{bardeen1973four}\cite{bekenstein1973black}. Although classical black holes were thought to be objects with zero temperature, from which no signal could escape, subsequent work argued that black holes should have an entropy proportional to the area of their horizons with proportionality constant 1/4 \cite{hawking1975particle}\cite{hawking1979euclidean}, this is the so-called Bekenstein-Hawking entropy:
\begin{equation}\label{eq:bekenstein-hawking}
S_{BH}=\frac{A_H}{4},
\end{equation}
where we consider natural units $G=c=\hbar=k_B=1$. Additional research \cite{gibbons1977cosmological}\cite{unruh1976notes}\cite{wald1993black} found that the thermodynamical behavior of black holes is just a particular case of a general feature associated with the presence of horizons in the coordinate chart under consideration. Therefore, it is possible to have thermodynamical phenomena in any spacetime provided that it contains a causally disconnected region. The converse statement, that thermodynamical relations can determine the geometry of the spacetime has been also studied, and it was found that this is indeed the case, provided certain assumptions. Remarkable works in this context include Jacobson's \cite{jacobson1995thermodynamics} and Padmanabhan's \cite{padmanabhan2010thermodynamical} approaches. In this context, the possibility that gravity is an emergent phenomenon that arises from the dynamics of a completely different quantum system has been explored, and many realizations have been discussed in literature such as Verlinde's entropic gravity proposal \cite{verlinde2011origin}. All these works have tried to obtain gravitational field equations from thermodynamics; however, in a more limited setting, it is possible to ask whether thermodynamical arguments can lead to specific solutions directly without considering Einstein equations. Some research has been devoted to such possibility, and it was found that spherically symmetric solutions can be obtained from thermodynamical relations for the internal energy understood as either the Misner-Sharp mass \cite{zhang2014schwarzschild}\cite{zhang2014thermodynamics} or Komar energy \cite{tan2017modified}. In both approaches for the relation of geometry with thermodynamics, it is clear that the identification of the thermodynamical variables is a matter of assumption.

In order to define and study the thermodynamical processes for black holes, one popular alternative is to identify the conserved charges of the spacetime as the relevant thermodynamical variables. This approach is well motivated in the context of usual thermodynamics since, in this framework, the observed quantities correspond to variables preserved by time averaging. As an example of this approach we can remark the classical work by Bardeen, Carter, and Hawking \cite{bardeen1973four}, in which the identification is based on the variation of the mass measured at infinity for Kerr-Newman black holes. Other examples in this context include \cite{gulin2017generalizations}\cite{ma2014corrected}\cite{barnich2005generalized}. Although this approach is well motivated, it has some issues related to the requirement of asymptotic flatness and, also, to negative heat capacities \cite{york1986black}. Another possibility is to derive relations from the Einstein equations that hold directly on the horizon and resemble thermodynamical laws. Such an approach has been studied in depth by Padmanabhan and coworkers (see for example \cite{padmanabhan2002classical}\cite{padmanabhan2010thermodynamical}). In this context, the identification of thermodynamical variables is based on the Bekenstein-Hawking analysis together with Einstein (or Lanczos-Lovelock \cite{kothawala2009thermodynamic}) equations for a perfect fluid to identify pressure and energy terms. Motivations based on the gravitational action do exist for this approach \cite{parattu2013structure}, and will be discussed later in the context of our results.

Finally, a possibility is that thermodynamical variables in gravitational systems are defined in terms of quantities constrained on their (finite) boundaries, this is the so-called quasilocal approach for thermodynamical quantities \cite{szabados2009quasi}. Evidently, this approach is independent of the global properties of the spacetime such as asymptotic flatness; however, the main issue with the quasilocal approach is that it is not evident which thermodynamical variable correspond to some combination of the quantities defined on the boundary. Examples within this approach are Hayward's approach \cite{hayward1998unified}, which considers the gradient of the Misner-Sharp energy by using the Einstein equations and identifies work terms according to the properties of matter and analogies with the special relativity case; and that due to Brown and York \cite{brown1993quasi}\cite{brown1993micro}. In this proposal, the identification of thermodynamical quantities is based on a Hamilton-Jacobi analysis of the on-shell Euclidean action for the gravitational partition function. A remarkable feature of this approach is that thermodynamic quantities are defined directly in terms of the canonical variables of the system, which is advantageous since it is evident how the Hamiltonian properties of a system determine a fundamental thermodynamical equation, from which all thermodynamics can be derived \cite{martinez1996fundamental}.

One thermodynamical relation of particular interest in the case of black holes is the Smarr relation. Its classical counterpart is the Euler equation \cite{callen1998thermodynamics}, a bilinear equation in extensive and intensive variables that illustrates the homogeneous character of thermodynamical variables. In the context of black hole thermodynamics, Smarr relations have additional numerical coefficients; for instance, for Kerr black holes of ADM mass $M$, angular momentum $J$ and electric charge $Q$   \cite{smarr1973mass}:
\begin{equation}\label{smarrkerr}
M=2TA_H+2\Omega J+\Phi Q,
\end{equation}
where the factors of 2 signal that the scaling properties of the Kerr spacetime are different from the usual thermodynamical systems, which in turn is interpreted as a consequence of the range and the absence of screening that characterize gravity. In the literature, Smarr relations are usually obtained through two ways: by using the Euler theorem for homogeneous functions in the first law of thermodynamics \cite{smarr1973mass}, and, on the other hand, by considering Komar-like integrals for the conserved charges (see for example \cite{gulin2017generalizations}\cite{ma2014corrected}\cite{kastor2009enthalpy}).

In the context of quasilocal variables, the analysis of the corresponding Smarr relations have been few in number. Discussions in this framework are typically abstract and centered in properties such as conservation, gauge-invariance, and positivity of mass. Although these analysis are important, practical applications of the quasilocal variables are interesting on their own \cite{lapierre2017cosmological}. Regarding black holes, there are approaches to obtain thermodynamical relations for quasilocal variables. The dominant point of view is based on Hayward's work \cite{hayward1998unified}. However, as we pointed out before, the identification of thermodynamical variables corresponding to matter is associated with analogies with the non-gravitational case, which exclude the possibility of identifying clearly quantities such as pressures and chemical potentials for the gravitational field itself. So, we are led to recognize that further exploration of quasilocal Smarr relations must be done.

In this work, we construct and analyze quasilocal Smarr relations for thermodynamical variables defined by using the Brown and York approach in the context of a static, spherically symmetric spacetime, considering the specific cases of absence of matter (Schwarzschild) and a spacetime with a electromagnetic field (Reissner-Nordstr\"om). As discussed before, the Hamilton-Jacobi approach provides a direct identification of thermodynamical variables and the fundamental equation of the system. Although this method has been applied to charged black holes in \cite{braden1990charged}, considering the thermodynamical stability conditions for this case and the first law, no Smarr relations were obtained. Regarding these relations in the context of Euclidean path integrals, an important recent work is \cite{banerjee2010statistical}, where the Smarr relation was obtained from the integrated action and some thermodynamical identities, although this work does not consider boundary terms and the convergence of the partition function is not guaranteed in this approach. Therefore, our work could be considered an improvement in this context.

Finally, we think that this work is relevant for the thermodynamical derivation of gravitational solutions. Previous research in this framework \cite{zhang2014schwarzschild}\cite{zhang2014thermodynamics} claim to be logically independent from Einstein field equations, but, being based on \cite{hayward1998unified}, they have included these equations in an implicit way. Our approach is suitable to specify what non-thermodynamic information is required to obtain solutions, given that it is expected on more general grounds \cite{padmanabhan2009entropy} that thermodynamical relations are not enough to recover the full form of Einstein equations without additional information.

This paper is organized as follows. In section \ref{sec:byquasilocal} we summarize the Brown-York method, to apply it to the spherically symmetric case in section \ref{sec:spherical}. The detailed study of the Maxwell-Einstein system is done in \ref{sec:rn}, where the Smarr relation is obtained. Finally, in section \ref{sec:discussion} we discuss the result and in section \ref{sec:finalremarks} we give some final remarks.

\section{Brown-York definition of quasilocal thermodynamical quantities}\label{sec:byquasilocal}
Usually, black hole thermodynamics and Smarr relations are described in terms of conserved charges defined at infinity; however, there are problems with a thermodynamical description of black holes in terms of these quantities. Schwarzschild black holes, for example, are understood as objects with a temperature $T_{BH}=\frac{1}{8\pi M}$, with $M$ the mass of the black hole. From thermodynamics we have that the heat capacity $C$ of a system satisfies:
\begin{equation}
C=-\beta^2\left(\ddpar{S}{E}\right)^{-1},
\end{equation}
where $S$ is the entropy and $E$ is the internal energy. Thus, taking into account that $\beta_{BH}=8\pi E=\dpar{S}{E}$, where we identified $E=M$, we have that a Schwarzschild black hole have a negative heat capacity. This result indicates that black holes are unstable as thermodynamical systems. There are two possible ways to circumvent this issue: to suppose that there is a negative cosmological constant in addition to the black hole (AdS-Schwarzschild black hole \cite{brown1994temperature}); or to consider that the black hole is contained in a bounded region with certain thermodynamical conditions fixed over the boundary, the so-called quasilocal approach. In this context, the thermodynamical system under consideration is defined as the spacetime inside the finite boundary, together with matter fields under consideration.

As a summary of the theoretical basis of our study we will review some aspects of Brown and York approach \cite{brown1993action}. In this framework we consider a spacetime $M$ that can be expressed as $\Sigma\times I$, with $I$ a real interval and $\Sigma$ a spacelike hypersurface. $\Sigma$ has a boundary $B$, and the product $B\times I$ is denoted $^3B$. $t'$ and $t''$ are the hypersurfaces associated to the endpoints of the interval (see Figure 1.).
\begin{figure}
\center{
\includegraphics[scale=0.5]{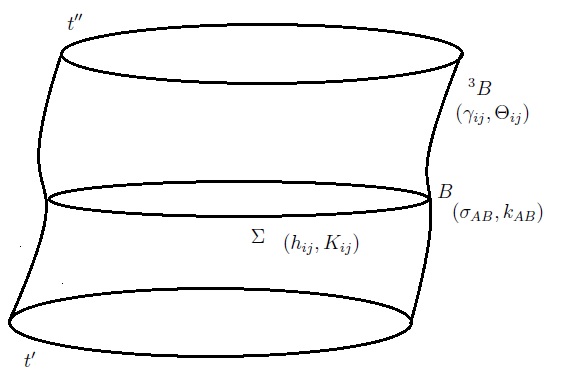}
\caption{Spacetime considered in the construction of quasilocal properties. $\Sigma$ and $^3B$ are spacelike and timelike 3-surfaces, respectively, and $B$ is a spacelike 2-surface. The notation for the corresponding metric and extrinsic curvature for each hypersurface is indicated in parenthesis. }
\label{fig1}}
\end{figure}\\
The gravitational action in $M$ in absence of matter is
\begin{eqnarray}
S^1=&&\frac{1}{16\pi}\int_Md^4x\sqrt{-g}{\it R}+\frac{1}{8\pi}\int_{t'}^{t''}d^3x\sqrt{h}K\nonumber\\
&&-\frac{1}{8\pi}\int_{ ^3B}d^3x\sqrt{-\gamma}\Theta,\label{eq:ehaction}
\end{eqnarray}
where $\gamma_{ij}$ and $\Theta$ correspond to the metric and extrinsic curvature (as embedded in $M$) of $^3B$, while $h$ and $K$ are the analogues for $\Sigma$. Boundary terms $S^0$ could be added to this action without altering the equations of motion. With an ADM decomposition for the metric $\gamma_{ij}$
\begin{equation}
\gamma_{ij}dx^idx^j=-N^2dt^2+\sigma_{ab}(dx^a+V^adt)(dx^b+V^bdt),
\end{equation}
The quasilocal stress energy tensor is constructed for the action $S=S^1+S^0$ as variations between classical solutions:
\begin{equation}
\tau^{ij}\equiv\frac{2}{\sqrt{-\gamma}}\frac{\delta S_{cl}}{\delta \gamma_{ij}}=\frac{2}{\sqrt{-\gamma}}(\pi^{ij}_{cl}-\pi^{ij}_{0}).
\end{equation}
Its projections are of special interest,
\begin{eqnarray}
\epsilon &&=u_iu_j\tau^{ij}=-\frac{1}{\sqrt{\sigma}}\frac{\delta S_{cl}}{\delta N},\\
j_a &&=-\sigma_{ai}u_j\tau^{ij},\\
s^{ab}&&=\sigma_i^a\sigma_j^b\tau^{ij},
\end{eqnarray}
since the variation of $S$ with respect to classical solutions is associated to the term on $^3B$ and given in terms of these quantities:
\begin{equation}
\delta S|_{^3B}=\int_{^3B}d^3x \sqrt{\sigma}\left(-\epsilon\delta N+j_a\delta V^a +\frac{N}{2}s^{sb}\delta\sigma_{ab}\right),
\end{equation}
where $\epsilon$, $j_a$, $s^{ab}$ are the quasilocal energy density, momentum density, and stress density, respectively. For static spacetimes $j_a=0$ and
\begin{equation}
\epsilon=\frac{1}{8\pi}k-\epsilon_0,
\end{equation}
\begin{equation}
s=\frac{1}{8\pi}\left(k^{ab}+(n_\mu a^\mu-k)\sigma^{ab}-(s_0)^{ab}\right).
\end{equation}
Where quantities with subindex 0 are due to $S^0$, $k^{ab}$ is the extrinsic curvature of $B$ as embedded in $\Sigma$ ($k$ is its trace), $n_\mu$ is the unit normal to $^3B$ and $a^\mu$ is the acceleration of $u_\mu$, the unit normal to $\Sigma$.

When describing a spacetime with a horizon in terms of these variables it is found that the variation of entropy is (see Appendix \cite{brown1993micro}):
\begin{eqnarray}
\delta S[\epsilon,j,\sigma]&&\approx\delta\left(\frac{A_H}{4}\right)\nonumber\\
&&=\int_Bd^2x\beta\left[\delta(\sqrt\sigma\epsilon)+\sqrt\sigma \frac{p^{ab}}{2}\delta\sigma_{ab}\right]\label{firstlaw}
\end{eqnarray}
Where $p^{}$ is defined in terms of time integrals of $s^{ab}$ and $N$. This expression resembles the first law of thermodynamics, and it will be the starting point for the analysis in what follows.
\section{Quasilocal analysis of Spherically symmetric spacetimes}\label{sec:spherical}
The specific case of spherical symmetry has been studied before \cite{brown1993action}, however we include the details of the calculations for future use. In spherically symmetric static spacetimes we consider the metric
\begin{equation}\label{metric}
ds^2=-N(r)^2dt^2+h(r)^2dr^2+r^2d\Omega^2.
\end{equation}
We must compute $\tau^{ij}$ for the metric (\ref{metric}). In this case, the hypersurfaces $\Sigma$ are surfaces defined by $t= constant$, with normal vector $u_\mu=-N\delta_\mu^0$. In addition, $^3B$ are hypersurfaces defined by $r=constant$ and with normal vector $n_\mu=h\delta^1_\mu$. We must note that this vector is defined outward the surface. With these prescriptions for the required hypersurfaces, the region $B$, on which the thermodynamical quantities are defined, is a $2-$sphere of radius $R$. Under these assumptions, and ignoring matter terms, a straightforward calculation leads to a vanishing extrinsic curvature for $\Sigma$:
\begin{equation}
K_{\mu\nu}=-h^\alpha_\mu\nabla_\alpha u_\nu=0,
\end{equation}
where $h_{\mu\nu}=g_{\mu\nu}+u_\mu u_\nu$ is the metric for $\Sigma$. With this, the canonical momentum associated to the $\Sigma$ hypersurfaces vanish:
\begin{equation}
P_{ij}=0.
\end{equation}
In the case of $^3B$, the metric for this surface is $\gamma_{\mu\nu}=g_{\mu\nu}-n_{\mu}n_\nu$. The corresponding extrinsic curvature is
\begin{eqnarray}
\Theta_{\mu\nu}&&=-\gamma^\alpha_\mu\nabla_\alpha n_\nu\nonumber\\
&&=-diag\left.\left(-\frac{N N'}{h},0,\frac{r}{h},\frac{r\sin^2\theta}{h}\right)\right|_{r=R}.
\end{eqnarray}
So
\begin{eqnarray}
\Theta&&=-\left.\left(\frac{N'}{Nh}+\frac{2}{rh}\right)\right|_{r=R}.
\end{eqnarray}
With these expressions, the canonical momentum associated to $^3B$ can be calculated:
\begin{widetext}
\begin{eqnarray}
\tau^{\mu\nu}_{cl}&&=-\frac{1}{16\pi}\sqrt{-\gamma}\left(\Theta \gamma^{\mu\nu}-\Theta^{\mu\nu}\right)\\
 &&=\frac{1}{8\pi}\left.diag\left(-\frac{2}{rN^2h}, 0, \frac{1}{r^2}\left(\frac{N'}{Nh}+\frac{1}{rh}\right), \frac{1}{r^2\sin^2\theta}\left(\frac{N'}{Nh}+\frac{1}{rh}\right)\right)\right|_{r=R}.
\end{eqnarray}
\end{widetext}
Finally, following the definitions and introducing the extra terms associated with the embedding of $^3B$ in a reference flat space, which essentially define the zero-point of energy \cite{brown1993quasi}:
\begin{equation}\label{energyd}
\epsilon=\left.\frac{1}{4\pi}\left(\frac{1}{r}-\frac{1}{rh}\right)\right|_{r=R},
\end{equation}
\begin{equation}\label{pressured}
p\equiv\frac12\sigma_{ab}s^{ab}=\left.\frac1{8\pi}\left(\frac{N'}{Nh}+\frac{1}{rh}-\frac{1}{r}\right)\right|_{r=R}.
\end{equation}
Defining the quasilocal energy as $E=\int_Bd^2x\sqrt\sigma\epsilon$ and restricting (\ref{firstlaw}) to this case it is obvious that
\begin{equation}\label{specific1stlaw}
T\delta S=\delta E+p\delta A.
\end{equation}
With $T=\frac{T_\infty}{N}$ the blueshifted temperature of the horizon. Now, following \cite{kastor2009enthalpy}, dimensional analysis tells us that $S$, $E$, $A$ scale with lenght as $(lenght)^2$, $(lenght)^1$, and $(lenght)^2$ respectively. Using the Euler theorem for homogeneous functions, which states that for an homogeneous function $f(x,y)$, satisfying $f(\alpha^p x,\alpha^q y)=\alpha^r f(x,y)$, the function and its derivatives satisfy:
\begin{equation}\label{eq:eulertheorem}
r f(x,y)=p\left(\dpar{f}{x}\right)x+q\left(\dpar{f}{y}\right)y,
\end{equation}
we have that (\ref{specific1stlaw}) implies that:
\begin{equation}\label{eq:vacuumsmarr}
2TS=E+2pA.
\end{equation}
Additional terms are to be expected when including matter fields, as in the Reissner-Nordstr\"om case shown below. We can see (\ref{eq:vacuumsmarr}) as the quasilocal Smarr relation for the vacuum black hole case. Replacing (\ref{energyd}) and (\ref{pressured}) it is found that
\begin{equation}
2TS=\frac1{N(R)}\left[\frac12 R^2\frac{(N(R)^2)'}{N(R)h(R)}\right]
\end{equation}
The factor in square brackets is precisely the Komar energy associated to the region considered, $E_K(R)$. Cancelling out the blueshift factors we have
\begin{equation}\label{result}
2T_\infty S=E_K(R).
\end{equation}
This expression agrees with previous results in the literature \cite{banerjee2010statistical}\cite{banerjee2010killing}, although there are some points to remark in this context. In \cite{banerjee2010statistical}, equation (\ref{result}) is obtained through a functional integral approach which does not consider boundary terms, as discussed before; in addition, the Komar energy is to be considered as the energy of the full spacetime. On the other hand, \cite{banerjee2010killing} shows that (\ref{result}) holds for a large class of static black hole spacetimes in $N+1$ dimensions when evaluated at the horizon radius. As we will see below, the presence of additional fields introduces additional terms in (\ref{result}) which vanish when $R$ is the horizon radius, in agreement with \cite{banerjee2010killing}.

To understand this result, it is convenient to consider the structure of the approach. The inputs of the equation are the expressions for the entropy, temperature and (vanishing) matter contribution. A relation with geometry is provided by the Komar charge since it involves the derivative of the metric, together with the expressions of temperature and entropy since they are associated with the metric and its derivative at the horizon.

Let us consider the Schwarzschild system of mass $M$, defined by $T_\infty=\frac{1}{8\pi M}$, $S=4\pi M^2$, and no matter fields. Two additional ingredients that we have not discussed explicitly are the Einstein field equations, which impose that $N(r)=h(r)^{-1}$, and the horizon radius.
With these elements, we see easily that (\ref{result}) is written as:
\begin{equation}
2\frac{1}{8\pi M}4\pi M^2=M=\frac12 R^2(N^2)'.
\end{equation}
This equation can be trivially integrated and gives $N(R)^2=1-\frac{2M}{R}$, as expected. This derivation of the Schwarzschild metric could be regarded as a thermodynamical approach to find solutions in the same sense as \cite{zhang2014schwarzschild}; however, the restriction $N(r)=h(r)^{-1}$ is not obtained through some thermodynamical condition so we consider this result as a realization of Einstein field equations in thermodynamical terms.

\section{Quasilocal Smarr relation for the Einstein-Maxwell system}\label{sec:rn}
Considerations for the Reissner-Nordstr\"om spacetime proceed along the same lines as the Schwarzschild case; however, we must take into account the contributions of the electromagnetic Hamiltonian to the boundary terms of the action. We start with the Maxwell Lagrangian:
\begin{equation}\label{eq:maxwelllagrangian}
L_M=-\frac{1}{16\pi}F_{\mu\nu}F^{\mu\nu},
\end{equation}
where
\begin{equation}
F_{\mu\nu}=A_{\nu;\mu}-A_{\mu;\nu},
\end{equation}
and $A^{\mu}=(\phi, \vec A)$ is the electromagnetic four potential. To perform the $3+1$ decomposition we consider a timelike surface-forming congruence and coordinates where the tangent vector field associated with the congruence is $t^\mu=\delta^\mu_0$. We consider spacelike slices $\Sigma$ such that $t^\mu=N n^\mu$, thus the shift functions vanish. In addition we consider a $S^2$ boundary for the slices. With this construction, the time derivatives of the field configuration variables have a simple form:
\begin{equation}
\dot{A}_\mu=\pounds_tA_\mu=A_{\mu;\nu}t^\nu+t_{\nu;\mu}A^\nu=A_{\mu;0}.
\end{equation}
In this context we have the metric
\[
   g_{\mu\nu}=
  \left( {\begin{array}{cc}
   -N^2 & 0 \\
   0 & h_{ij} \\
  \end{array} } \right).
\]
Now, the canonically conjugated momenta for this system are:
\begin{equation}\label{eq:momenta}
p^\mu=\dpar{}{\dot{A}_\mu}\left(\sqrt{-g}\mathcal{L}_M\right)=-\frac{\sqrt{-g}}{4\pi}F^{0\mu}.
\end{equation}
Antisymmetry of $F_{\mu\nu}$ implies that $p^0=0$, so we have a constraint. Inverting (\ref{eq:momenta}) for the non vanishing momenta we find that
\begin{equation}
A_{i;0}=4\pi\frac{N}{\sqrt{h}}h_{ij}p^j+A_{0;i}
\end{equation}
With this relation, the Hamiltonian density is
\begin{equation}\label{eq:hamiltoniandensity}
\mathcal{H}_M=p^i\left(4\pi\frac{N}{\sqrt{h}}h_{ij}p^j+A_{0;i}\right)-\mathcal{L}_M N\sqrt{h}.
\end{equation}
After some algebra, it is found that
\begin{eqnarray}
\frac{\mathcal{H}_M}{\sqrt{h}}=&&4\pi\frac{N}{h}p_ip^i-\left(\frac{N}{4\pi}A_0F^{0i}\right)_{;i}\nonumber\\
&&+A_0\left(\frac{N}{4\pi}F^{0i}\right) _{;i}-N\mathcal{L}_M.\label{hamiltonianscalar}
\end{eqnarray}
From this expression and the vanishing of $p^0$ we have that $A_0$ is a Lagrange multiplier that implements the Gauss law constraint $(N F^{0i})_{;i}=0$. For our purposes it is important to consider the second term of the right hand side of (\ref{hamiltonianscalar}), this term is a divergence that produces a boundary term in the action. We have that the electromagnetic term of the action is
\begin{widetext}
\begin{eqnarray} %\nonumber % Remove numbering (before each equation)
  S_m =&&\int dt\left\{\int_{\Sigma_t}d^3y\sqrt{h}\left[\frac{p^iA_{i;0}}{\sqrt{h}}-4\pi\frac{N}{h}p_ip^i-A_0\left(\frac{N}{4\pi}F^{0i}\right)_{;i}+N\mathcal{L}_M\right]\right. \nonumber\\
   &&\left.+\oint_{B_t}\frac{1}{4\pi}NA_0\left[-\frac{1}{N^2}h^{ij}F_{0j}\right]r_i\sqrt{\sigma}d^2\theta\right\}
\end{eqnarray}
\end{widetext}
This canonical expression is useful to identify the quasilocal quantities that can be derived from its variation; however, regarding the first law of thermodynamics, the variation of the action based on the Lagrangian (\ref{eq:maxwelllagrangian}) could involve the variation of non-canonical variables, in such a way that the obtained action would not correspond to the microcanonical ensemble, leading to an incorrect identification of the thermodynamical variables of the first law.
\par To remedy this, it is useful to consider the approach by Iyer and Wald \cite{creighton1996gravitational}\cite{iyer1994some}, in which black hole entropy can be obtained by considering the  Noether current associated to diffeomorphism invariance. Regarding the Brown and York approach, it has been shown that the two approaches are equivalent and that the microcanonical action can be obtained through a boundary term constructed from the Noether charge \cite{iyer1995comparison}. To implement the Noether charge approach, we need to consider the Lagrangian version of the action and identify the fields under variation.
The Lagrangian tensor density is given by
\begin{equation}\label{eq:maxwelldensity}
\mathcal{L}_M=\sqrt{-g}L_M=-\frac{1}{16\pi}\sqrt{-g}F_{\mu\nu}F^{\mu\nu}.
\end{equation}
The variation of this density is:
\begin{equation}
\delta\mathcal{L}_M=\frac{1}{16\pi}\delta(\sqrt{-g}g^{\mu\rho}g^{\nu\sigma}F_{\rho\sigma}F_{\mu\nu}).
\end{equation}
After some algebra
\begin{eqnarray}
\delta\mathcal{L}_M=&&-\frac{\sqrt{-g}}{16\pi}\left[\left(2F_\alpha^{\ \nu}F_{\nu\beta}-\frac12g_{\alpha\beta}F^{\mu\nu}F_{\mu\nu}\right)\delta g^{\alpha\beta}\right.\nonumber\\ &&\left.+\left(-4F^{\rho\sigma}_{\phantom{\rho\sigma};\rho}\right)\delta A_\sigma+\left(4F^{\rho\sigma}\delta A_\sigma\right)_{;\rho}\right].
\end{eqnarray}
The last term of the right hand side corresponds, when considered in the action integral, to an exact differential form \boldmath${d \rho}$\unboldmath, with components
\begin{equation}\label{eq:rhoform}
\rho^\alpha=-\frac{1}{4\pi}F^{\alpha\sigma}\delta A_\sigma
\end{equation}
This object will be relevant for the construction of the Noether charge. From the previous results, the variation of the matter action $S_M$ is
\begin{equation}
\delta S_M=\int_{\mathcal{M}}d^4x\sqrt{-g}\left(-\frac12 T_{\alpha\beta}\delta g^{\alpha\beta}+(E_A)^\sigma \delta A_\sigma+\rho^\alpha_{\phantom{\alpha};\alpha}\right),
\end{equation}
where
\begin{eqnarray}
T_{\alpha\beta}&&=\frac{1}{8\pi}\left(2F_\alpha^{\ \nu}F_{\nu\beta}-\frac12g_{\alpha\beta}F^{\mu\nu}F_{\mu\nu}\right),\\
(E_A)^\sigma &&=-4F^{\rho\sigma}_{\phantom{\rho\sigma};\rho},
\end{eqnarray}
are the factors associated with the equations of motion; therefore, the on-shell action will involve only the $\rho^\alpha$ term, which thus will be associated with the thermodynamics. The Noether current associated with a diffeomorphism generated by a vector field $\xi^a$ is
\boldmath
\begin{equation}\label{eq:noethercurrent}
j[\xi]=\rho_\xi-\xi\cdot\mathcal{L}_M,
\end{equation}
\unboldmath
where $\rho_\xi$ indicates that all variations in $\rho$ are to be replaced by Lie derivatives along $\xi^a$. Explicit evaluation of this expression gives
\begin{equation}
j^a[\xi]=-\xi_bT^{ab}-(E_A)^a\xi^b A_b-\left(F^{ab}\xi^c A_c\right)_{;b}.
\end{equation}
When the equations of motion hold, this current corresponds to a total divergence that defines the conserved Noether charge:
\begin{equation}\label{eq:ncharge}
q[\xi]=\Xi\xi^bA_b,
\end{equation}
where
\begin{equation}\label{eq:Xi}
\Xi=\frac{1}{4\pi}r_bE^b,
\end{equation}
with $r_b$ the spacelike normal to $B$. The Iyer and Wald approach shows that the microcanonical action $I_{mic}$ can be written as
\begin{equation}\label{eq:microcanonical}
S_{mic}=\int_{\mathcal{M}}d^4x\sqrt{-g}\mathcal{L}_M-\int_{^3B}d^3x\sqrt{\gamma}q[\xi].
\end{equation}
When considering diffeomorphisms along $t$, we have that
\begin{equation}
q[t]=N\Phi\Xi.
\end{equation}
With this, the variation of the electromagnetic part of the microcanonical action is simply given by
\begin{equation}\label{eq:microvariation}
\delta S_{mic}=\int dt\int_{B}d^2\theta\sqrt{\sigma} N\Phi\delta\Xi,
\end{equation}
thus, there is no electromagnetic contribution to quasilocal energy, momentum, and pressure. The effect of electromagnetism is to include an additional work term, that can be interpreted as a chemical potential after noting that $\Xi$ can be regarded as a superficial charge density because of the Gauss law.
The first law of thermodynamics for the static Einstein-Maxwell system is, therefore:
\begin{equation}
 \delta S=\int_{B}d^2\theta\sqrt{\sigma}\beta\left(\delta\epsilon+s^{AB}\delta\sigma_{AB}+\Phi\delta\Xi \right)
\end{equation}

\subsection{Derivation of the Smarr Relation}
In the spherically symmetric case we have that the only non-vanishing component of the electromagnetic tensor is
\begin{equation}
F_{tr}=\frac{Q}{r^2}
\end{equation}
With this, the electrostatic potential can be obtained by integration,
\begin{equation}
\Phi(r)=\frac{Q}{N(r)}\left(\frac{1}{r}-\frac{1}{r_H}\right),
\end{equation}
where the requirement of a finite potential on the horizon has been considered \cite{gibbons1977action} to fix the integration constant.
In addition, integration of the first law on the spherical quasilocal surface or radius $r$ leads to
\begin{equation}
T\delta S=\delta E+ s\delta A+\Phi\delta Q.
\end{equation}
Using Euler's theorem, the quasilocal Smarr relation for Reissner-Nordstr\"om black holes  is obtained:
\begin{equation}\label{eq:RNquasilocalSmarr}
2TS=E+2sA+\Phi Q.
\end{equation}
The corresponding expressions for the thermodynamical variables in this case are written as:
\begin{eqnarray}
E &&=-r\frac{1}{h(r)},\\
s &&=\frac{1}{8\pi h(r)}\left(\frac{1}{r}+\frac{d}{dr}[\log N(r)]\right),\\
T &&=\frac{1}{N(r)}\frac{2-2M/r_H}{4\pi r_H},\\
A &&=4\pi r^2,\\
S &&=\frac{A_H}{4}=\pi r_H^2.
\end{eqnarray}
If we introduce the expressions for the quasilocal energy and pressure, and electric potential in (\ref{eq:RNquasilocalSmarr}), and follow the same procedure as in the vacuum case, we obtain that (\ref{result}) contains an additional term:
\begin{equation}\label{eq:Komarandpotential}
2T_\infty S=  E_K(R)+Q^2\left(\frac{1}{R}-\frac{1}{r_H}\right).
\end{equation}
If we let $R=r_H$, the electrostatic potential term vanishes and we recover the result of \cite{banerjee2010killing}.

Inserting the full set of thermodynamical variables into (\ref{eq:RNquasilocalSmarr}) we obtain, after some cancellations,
\begin{equation}
\left(r_H-M+\frac{Q^2}{r_H}\right)-\left(\frac{Q^2}{r}\right)=N(r)\frac{r^2}{h(r)}\left(\frac{d}{dr}[\log N(r)]\right).
\end{equation}
By using the horizon condition that defines $r_H$, namely $r_H^2-2Mr_H+Q^2=0$, we have
\begin{equation}
M-\left(\frac{Q^2}{r}\right)=N(r)\frac{r^2}{h(r)}\left(\frac{d}{dr}[\log N(r)]\right).
\end{equation}
Einstein equations for the electrovacuum system imply that $N(r)h(r)=1$, therefore:
\begin{equation}
M-\frac{Q^2}{r}=r^2N(r)\frac{dN(r)}{dr}.
\end{equation}
\begin{equation}
\frac{2M}{r^2}-\frac{2Q^2}{r^3}=\frac{d}{dr}[N(r)^2].
\end{equation}
This equation can be integrated trivially to give
\begin{equation}
N^2(r)=1-\frac{2M}{r}+\frac{Q^2}{r^2},
\end{equation}
as expected.

\section{Discussion}\label{sec:discussion}
There are some important points to discuss regarding the results presented in previous sections. In the following, we will comment the relations of our approach and our results with the other frameworks mentioned before for black hole thermodynamics. In addition, we identify and discuss possibilities for further research.

A crucial difference between Hamilton-Jacobi methods, used by us, and Hayward's approach is the presence of a pressure associated with the gravitational field in the first one. According to the relation that we found between the quasilocal variables of Brown and York and the Komar energy, which is related to the Misner-Sharp energy through a Legendre transformation \cite{tan2017modified}\cite{chen2011first}, we can conclude that this additional pressure is included in the thermodynamical potential that plays the role of internal energy in Hayward's work. Consequently, in the context of Brown and York variables, Komar and Misner-Sharp energies could be understood as enthalpies where the finite Legendre transform has a numerical factor related with the scaling properties of the variables. Usually, pressure variables such as the pressure of the cosmological constant are considered as volumetric pressures, and a conjugated thermodynamical volume can be constructed accordingly \cite{kubizvnak2017black}; these quantities are related to the quasilocal areal variables by virtue of the spherical symmetry, so they can be compared with the results of our approach.

\par In addition, it is interesting to compare the features of horizon thermodynamics and Hamilton-Jacobi methods for the description of gravitational thermodynamics. First of all, it is interesting to remark that Hamilton-Jacobi approaches rely on boundary terms for specific foliations to define the thermodynamical quantities, whereas horizon thermodynamics does not introduce such terms, but it considers only the Einstein-Hilbert action \cite{parattu2013structure}; this is convenient since in this form the gravitational action is holographic in the sense that boundary and bulk terms are related. However, from the point of view of Brown and York, the absence of a boundary term is still an election for the boundary term that is equivalent to suppose a microcanonical ensemble. In this sense, it is interesting to note that the projections of the holographically conjugated variables $f^{\mu\nu}=\sqrt{-g}g^{\mu\nu}$ and $N^\rho_{\mu\nu}$, introduced by \cite{parattu2013structure}, on our particular foliation have a form similar to the quantities appearing in (\ref{firstlaw}). This is to be expected since the variational principle associated with these variables fixes the momenta on the boundary and the quasilocal energy and momentum densities that appear in the quasilocal first law are constructed from the same momenta. In \cite{parattu2013structure} it is noted that holographic canonical conjugacy does not imply that the integral surface Hamiltonian gives the entropy, we consider this result to be related to the preceding remark, because different sets of canonically conjugated variables are expected to be related by an extended canonical transformation, and this is implemented in the gravitational setting by including boundary terms which correspond to Legendre transformations to other statistical ensembles, with different thermodynamical potentials. Thus, it is reasonable to expect that a large set of possible conjugated variables will not lead to entropy but to other potentials. The explicit implementation of this idea could be an interesting question for additional research. With respect to the thermodynamical variables, it must be noted that thermodynamical variables are identified in horizon thermodynamics from Einstein (or Lanczos-Lovelock) field equations for a perfect fluid; again, it is difficult to associate energy or pressure to the gravitational field in this context. Energy densities and pressures are local consequently, and integrated in a volume, which has issues when applied to black holes with a singularity. Finally, we must remark that the corresponding expression for our results in horizon thermodynamics is
\begin{equation}\label{eq:smarrht}
E_K=2TS.
\end{equation}
As shown in \cite{banerjee2010killing}, the validity of this result extends beyond our setting and is independent of foliations, so it can be interpreted as a property of horizons. This interpretation leads to the equipartition principle proposed by Padmanabhan \cite{padmanabhan2010equipartition}, which allows a heuristic understanding of the microscopic degrees of freedom that lead to gravitation \cite{vargas2018sads}. Our results can be regarded as an extension of (\ref{eq:smarrht}) for general radii $r$, and it is natural to ask whether the additional terms imply modifications to the equipartition proposal. This could be explored in future work.

Regarding the Smarr relation (\ref{eq:RNquasilocalSmarr}), the presence of terms evaluated at the horizon agrees with other results \cite{gulin2017generalizations}. These terms are important since they cancel out the term asociated with horizon temperature and entropy in such a way that a reminiscent of Einstein equations is obtained; however, it must be noted that we do not claim that Einstein field equations are replaced by our procedure, since the condition $N(r)h(r)=1$ must be introduced to recover the metric function, In addition, the equation for the horizon must be considered in advance. So, it is important to state that our results are not an alternative to the field equations, but a thermodynamical realization in which they are implicit. The same observation could be made for other works in this context such as \cite{zhang2014schwarzschild}\cite{tan2017modified}. Specifically, these works consider metrics of the form
\begin{equation}\label{schwarzschildgauge}
ds^2=-f(r)dt^2+\frac{1}{f(r)}dr^2+r^2d\Omega^2
\end{equation}
which is not the most general spherically symmetric metric but a particular case in which the condition that we needed to supplement ($N(r)h(r)=1$ in our notation) is supposed from the very beginning. As discussed in \cite{poisson2004relativist}, a condition must be fulfilled by the stress-energy tensor
\begin{equation}
-T^t_t+T^r_r=0,
\end{equation}
in order to express the metric in the form (\ref{schwarzschildgauge}). This condition is due to Einstein field equations, and it could have an analogous form for other theories of gravity. The important point is that such conditions are implicit when metrics like (\ref{schwarzschildgauge}) are considered, therefore some information respecting the gravitational field equations is implicit in the thermodynamical derivation, even when it is claimed that this is not the case. Padmanabhan has argued \cite{padmanabhan2002essay} that it is possible to recover the general relativity action, and thus Einstein equations, from the Bekenstein-Hawking entropy, but some non-thermodynamical information must be supplemented such as the principle of general covariance. We regard this fact as an indication that our result on the necessity of supposing $N(r)h(r)=1$ in addition to thermodynamics to recover the Einstein equations is well founded.

It must be noted that our approach could provide thermodynamical restrictions on the possible sources of gravity allowed by Einstein equations. Let us consider a simple example: an AdS spacetime described by
\begin{equation}\label{eq:ads}
ds^2=-\left(1+\frac{r^2}{l^2}\right)dt^2+\left(1+\frac{r^2}{l^2}\right)^{-1}dr^2+r^2d\Omega^2.
\end{equation}
This spacetime is trivial in the sense that it does not have horizons and its Euclidean section is simply connected; with these features, there is no gravitational entropy, and no matter contribution to this quantity is to be expected since the entropy density $s_m=\frac{p+\rho}{T}$ for matter vanishes. Therefore, we expect that the quasilocal Smarr relation have the form:
\begin{equation}
0=E+2pA+\mbox{matter contributions},
\end{equation}
where the matter contributions arise from the boundary terms of the matter Lagrangian.
If the cosmological constant is implemented as a geometric property of the spacetime, it must correspond to a constant term $-2\sqrt{-g}\Lambda$ in the gravitational action. The variation of this term only involves changes in the metric, thus there is no boundary contribution in this case. Therefore, the quasilocal Smarr relation has no matter contribution:
\begin{equation}
0=E+2pA.
\end{equation}
However, by using (\ref{eq:ads}) in (\ref{energyd}) and (\ref{pressured}) it is obtained that
\begin{equation}\label{eq:mismatch}
E+2pA=\frac{R^3}{l^2}\left(1+\frac{R^2}{l^2}\right)^{-\frac12}\neq 0.
\end{equation}
To get a better understanding of this result, let us briefly discuss the results of this analysis for the AdS spacetime with a cosmological constant implemented by a slow-rolling scalar field $\phi$. The matter Lagrangian is
\begin{equation}\label{eq:scalarlagrangian}
L_M=-\frac12g^{\mu\nu}\nabla_\mu\phi\nabla_\nu\phi - V(\phi).
\end{equation}
The cosmological constant is recovered by considering a solution in which $\nabla\phi$ is negligible, with which $\rho=V(\phi)=-p$. However, it can be shown that the boundary contribution of the variation of the action is proportional to $(\nabla_\mu\phi)\delta\phi$, which vanishes in the slow-roll regime. Therefore the result (\ref{eq:mismatch}) is recovered also in this setting. This can be understood as a quasilocal realization of the thermodynamical issues of dark energy fluids that have been studied in the literature \cite{barboza2015thermodynamic}. Given these caveats, more complete analysis of the implications of these results for the de Sitter and Anti de Sitter cases are needed.

With respect to the rotating stationary black hole case, some comments are in order. Brown and York formalism is quite general and it includes the possibility of rotating stationary spacetimes; in fact, the general form of the quasilocal first law contains an angular momentum term (see Eq. (\ref{eq:generalfirstlaw}) in the Appendix). In this context, calculations have been done for specific rotating black hole metrics and different choices of boundary two$-$surfaces and reference terms; however, the research has been focused on studying the properties of the quasilocal energy for rotating spacetimes to establish its consequences as a measure of energy. Some examples of these studies include \cite{martinez1994quasilocal}\cite{maluf1996gravitational}\cite{schmekel2018quasilocal}. In the recent work \cite{schmekel2018quasilocal}, full expressions are given for the quasilocal stress-energy tensor assuming a null reference term, and it is clear that rotating metrics are technically challenging to study and the corresponding expressions include a non-trivial angular dependence which makes difficult the integration. However, regarding the quasilocal Smarr relation, the main issue to extend the results shown in this work to the rotating black hole case is related with a more fundamental property of the Kerr-Newman metric, namely, the impossibility to simultaneously set constant values on a specific surface $B$ for $\beta$ and the chemical potentials for angular momentum and other conserved charges \cite{brown1991complex}. This issue implies that the quasilocal first law (\ref{eq:generalfirstlaw}) can not be easily integrated on $B$ to obtain a bilinear form in conjugated extensive and intensive variables through homogeneity arguments; therefore, there is no quasilocal Smarr relation in a simple form as in the static cases considered above, but it has a convoluted form that resembles, for example, the corresponding expression in black holes coupled with nonlinear electrodynamics \cite{gulin2017generalizations}. Usual derivations of the Smarr relation for rotating black holes \cite{smarr1973mass} are not affected by this problem since the surfaces of constant $\beta$ and chemical potentials coincide at infinity. It remains to be studied whether the resulting Smarr$-$like expression for rotating black holes can be simplified as shown in reference \cite{gulin2017generalizations} for nonlinear electrodynamics, but this is beyond the scope of this work.

The results here presented could be interesting in the context of emergent gravity, in the same way that the thermodynamics of horizons motivated many theoretical developments in this sense \cite{padmanabhan2010thermodynamical}. For example, we obtained a fundamental difference with the entropic force approach \cite{verlinde2011origin}\cite{chen2011first} regarding the temperature of the 2-boundaries of the system, the so-called screens in this setting. In such approach, temperature is defined as an Unruh temperature (proportional to the radial derivative of the metric at the point under consideration), whereas we consider the blueshifted temperature of the horizon, which can be more easily understood physically and fulfills the role that temperatures should play in the partition function as seen from the Euclidean path integral. Additionally, as discussed before, we see that Komar energy is not an internal energy in a fundamental sense, therefore any approach that wants to consider gravity as an emergent effect from another theory and is based in a Lagrangian variational principle, should obtain the quasilocal Brown-York energy (or other Hamiltonian-based proposals) through statistical mechanics in such a framework. The specific implementation of this idea in a concrete situation will be considered in future work.

\section{Final Remarks}\label{sec:finalremarks}
We obtained Smarr relations not only valid for horizon quantities but that involve quantities like the quasilocal energy, defined at the finite boundary of a spherically symmetric region. The approach we used to obtain quasilocal Smarr relations is general from an operational point of view and can be summarized as follows: quasilocal quantities for the considered boundary must be found, including possible contributions from the matter Hamiltonian. Afterwards, the resulting expressions should be inserted in the first law, whose form depends on the topology of the considered manifold and the presence of horizons \cite{brown1993micro}. Finally, Euler theorem for homogeneous functions must be used to obtain the quasilocal Smarr relation. The homogeneous property for gravitational variables is expected to be valid in general \cite{padmanabhan2010gravitation}. Thus, we think that this construction can be generalized to other approaches for quasilocal variables based on Hamilton-Jacobi analysis of the action such as \cite{wang2009quasilocal}\cite{hollands2005comparison}.

\section{Acknowledgments}
P. B. acknowledges the support from the School of Sciences and Vicerrector\'{\i}a de Investigaciones of
Universidad de Los Andes, Bogot\'a, Colombia. F. V. acknowledges the support from the School of Sciences of Universidad de los Andes.

\appendix
\section{Derivation of the quasilocal first law}
The quasilocal first law of thermodynamics, Eq. (\ref{firstlaw}), which is an important property of quasilocal variables, underlies our approach. For the sake of completeness, in this Appendix we summarize the procedure given in \cite{brown1993micro} to obtain this relation.

Usual derivations of the first law of black hole thermodynamics are based on geometrical relations supplemented by Einstein field equations and energy conditions. However, Brown and York approach starts from the evaluation of the partition function through Euclidean path integrals in the same way as Gibbons and Hawking \cite{gibbons1977action}, although the derivation considers a spatially finite region and does not take the spacelike infinity limit. In this context, quantities defined on the boundary of the region under consideration define the statistical ensemble considered for the system.
The calculation starts with the Einstein-Hilbert action together with the Gibbons-Hawking-York (GHY) term, Eq. (\ref{eq:ehaction}):
\begin{eqnarray}
S^1&&=\frac{1}{16\pi}\int_M d^4x \sqrt{-g}R+\frac{1}{8\pi}\int_{t'}^{t''}d^3x\sqrt{h}K\nonumber\\
&&-\frac{1}{8\pi}\int_{^3B}d^3x\sqrt{-\gamma}\Theta.\label{actionwithghy}
\end{eqnarray}
It is widely known that the GHY term is introduced to obtain a well defined variational principle for gravitation in which metric components of the metric on the boundary are fixed, which implies that the variation of the action is:
\begin{eqnarray}
\delta S^1&&=(\mbox{e. o. m. terms})+\int_{^3B}d^3x\sqrt{\sigma}\left(-\epsilon\delta N\right.\nonumber\\
&&\left.+j_a\delta V^a+\frac{N}{2}s^{ab}\delta\sigma_{ab}\right),\label{eq:appxvariation}
\end{eqnarray}
where terms on $t'$ and $t''$ have been discarded since they coincide in the construction of the path integral for the partition function as a trace so they cancel out.

An essential point is to note that, in the path-integral formalism, temperature is related with the invariant lenght of the integral in the imaginary time coordinate, $\beta=\int d\tau N$, therefore, to fix $N$ is equivalent to fix a temperature for the partition function. This implies that the partition function based on the action (\ref{actionwithghy}) corresponds to a (grand) canonical ensemble since $N$ is fixed in the variational principle. Brown and York discuss \cite{brown1993micro} that it is necessary to consider a microcanonical ensemble to construct a density of states, which leads directly to entropy. Formally, the density of states can be found from the canonical partition function by performing a Laplace transform (which requires certain conditions to be well defined). This transform is equivalent, for the path integral, to the addition of a boundary term to the action (\ref{actionwithghy}), in the same way that to add a total derivative to the Lagrangian changes the variables to be fixed at the boundary when constructing the variational principle.

The extensive variables of a system can be defined as those constructed in terms of canonical variables, so that the microcanonical ensemble is defined as the ensemble of systems on which those variables are fixed on the boundary. The microcanonical action is obtained by adding the appropriate term:
\begin{eqnarray}
% \nonumber % Remove numbering (before each equation)
  S_m &=& S^1+\int_{^3B}d^3x\sqrt{\sigma}\left(N\epsilon-V^aj_a\right) \\
   &=& \int_M d^4x\left(P^{ij}\dot{h}_{ij}-N\mathcal{H}-V^i\mathcal{H}_i\right),
\end{eqnarray}
where the Hamiltonian form of the action with canonical coordinates $h_{ij}$ and momenta $P^{ij}$ is considered, and $\mathcal{H}$ and $\mathcal{H}_i$ are the energy and momentum constraints. The variation of $S_m$ can be written as:
\begin{eqnarray}
% \nonumber % Remove numbering (before each equation)
  \delta S_m &=& \mbox{(e. o. m. terms)}+\int_{^3B}d^3x\left[N\delta(\sqrt{\sigma\epsilon})\right.\nonumber\\
  &&\left.-V^a\delta(\sqrt{\sigma}j_a) +\frac{N\sqrt{\sigma}}{2}s^{ab}\delta\sigma_{ab}\right].\label{eq:microcanvariation}
\end{eqnarray}
Evidently, variables $\sqrt{\sigma}\epsilon$, $\sqrt{\sigma}j_a$, and $\sigma_{ab}$ are constructed from the canonical variables, so they can be regarded as extensive variables. Therefore, variation (\ref{eq:microcanvariation}), when considered on-shell, shows why $S_m$ can be considered as the microcanonical action for this system, where extensive variables are fixed on the boundary.

Formally, the density of states can be written as a path integral where $\epsilon$, $j$, and $\sigma$ are held fixed on the boundary
\begin{equation}\label{eq:densitypathint}
\nu[\epsilon,j,\sigma]=\sum_M \int \mathcal{D}H\exp{(iS_m)},
\end{equation}
where the sum over manifolds $M$ includes different topologies that respect the requirement of a $B\times S^1$ topology for the boundary. In the case of black holes, it is necessary to consider an axisymmetric solution where the stationary time slices contain the closed orbits of the axial Killing vector field. In addition, let $B$ be a topologically spherical two$-$surface that contains the orbits of the axial Killing vector field, and is contained
in a constant time hypersurface. It is important to note that the Lorentzian black hole metric can not be an extremum of $S_m$ since it does not have a $S^2\times S^1$ boundary; however, the complex metric constructed by the imaginary time prescription $T\rightarrow -iT$ fulfills this requirement \cite{brown1993micro}. Let the Lorentzian black hole metric be written as
\begin{equation}\label{eq:bhmetric}
ds^2=-N^2dT^2+h_{ij}(dx^i+V^idT)(dx^j+V^jdT),
\end{equation}
where the metric functions $N$, $h_{ij}$, and $V^i$ are time-independent. The horizon corresponds to $N=0$, and let coordinates be chosen in such a way that $V^i/N=0$ at the horizon. With the replacement $T\rightarrow -iT$, the complex black hole metric is obtained
\begin{equation}\label{eq:complexbh}
ds^2=-(-iN)^2dT^2+h_{ij}(dx^i-iV^idT)(dx^j-iV^jdT),
\end{equation}
where $T$ is real. The $N=0$ two$-$surface is called the bolt. In order to obey Einstein equations at the bolt, conical singularities must be avoided at the bolt. This condition is equivalent to
\begin{equation}\label{eq:boltcondition}
P(n^i\partial_i N)=2\pi,
\end{equation}
where $P$ is the period of coordinate $T$ and $n^i$ is the spacelike normal to the bolt. This condition defines the inverse temperature of the black hole through $P=2\pi/\kappa_H$, where $\kappa_H$ is the surface gravity. The complex metric (\ref{eq:complexbh}) together with (\ref{eq:boltcondition}) extremizes $S_m$ and can be used to construct a steepest descent approximation to the density of states:
\begin{equation}\label{eq:steepest}
\nu[\epsilon,j,\sigma]\approx\exp(iS_m[-iN,-iV,h]),
\end{equation}
with $S_m[-iN,-iV,h]$ the microcanonical action evaluated at the complex extremum. In addition, the density of states is related with the entropy $S$ as
\begin{equation}\label{eq:entropyandnu}
\nu[\epsilon,j,\sigma]=\exp(S[\epsilon,j,\sigma]),
\end{equation}
Thus,
\begin{equation}\label{eq:entropyasaction}
S[\epsilon,j,\sigma]\approx iS_m[-iN,-iV,h]
\end{equation}
Replacing this in the on-shell version of (\ref{eq:microcanvariation}), cancelling out appropriately the imaginary factors, and considering the expression for the inverse temperature stated before, it is obtained that
\begin{eqnarray}
\delta S[\epsilon,j,\sigma]=&&\int_B d^2x\beta\left[\delta(\sqrt\sigma \epsilon)-\beta\omega\delta(\sqrt{\sigma}j_a\sigma^a)\right.\nonumber\\
&&\left.+\beta\frac{\sqrt{\sigma s^{ab}}}{2}\delta\sigma_{ab}\right].\label{eq:generalfirstlaw}
\end{eqnarray}
Which is the general version of the quasilocal first law (\ref{firstlaw})
\bibliography{Bibliography}

\end{document}